\documentstyle[preprint,aps]{revtex}
\title{Hamiltonian approach to the transport properties 
of superconducting quantum point contacts}
\author{J. C. Cuevas, A. Mart\'{\i}n-Rodero and A. Levy Yeyati }
\address{
Departamento de F\'\i sica Te\'orica de la Materia Condensada C-V.\\
Facultad de Ciencias. Universidad Aut\'onoma de Madrid.\\
E-28049 Madrid. Spain.}
\begin{document}

\draft
\maketitle

\begin{abstract}
A microscopic theory of the transport properties of 
quantum point contacts giving a unified description of the
normal conductor-superconductor (N-S) and superconductor-superconductor (S-S)
cases is presented. It is based on a
model Hamiltonian describing   
charge transfer processes in the contact region and makes use of
non-equilibrium Green function techniques for the calculation of
the relevant quantities. It is explicitly shown that
when calculations are performed up to infinite order in the
coupling between the electrodes,
the theory contains all known results
predicted by the more usual scattering approach for N-S and S-S
contacts. For the latter we introduce a specific
formulation for dealing with the non-stationary transport properties. 
An efficient algorithm is developed for obtaining the dc and
ac current components, which allows a detailed analysis of the
different current-voltage characteristics for all range of parameters. 
We finally address the less understood small bias limit, for
which some analytical results can be obtained within the
present formalism. It is shown 
that four different physical regimes can be reached in this limit
depending on the values of the inelastic scattering rate and the
contact transmission. The behavior of the system in these regimes 
is discussed together with the conditions for their experimental
observability.
\end{abstract}

PACS numbers: 74.50.+r, 85.25.Cp, 73.20.Dx
\vspace{1cm}

\narrowtext

\section{Introduction}

Electronic transport through N-S and S-S junctions and weak links
has been the object of interest for many years \cite{general,Likharev}. 
Much of the theoretical understanding of these systems 
has been obtained by analyzing simple models where a one-dimensional
character is assumed \cite{Likharev,BTK}. These kind of models have been very
useful for clarifying the microscopic origin of basic phenomena like
the excess current in N-S and S-S contacts and the subgap structure in
S-S and S-Insulator-S (S-I-S) junctions. 
In particular, the one-dimensional scattering model
introduced by Blonder et al. \cite{BTK}
(hereafter referred to as BTK) has provided
a simple way of analyzing the transport properties in terms of normal
electron transmission and Andreev reflection probabilities. 

With the recent advances in the fabrication of nanoscale devices
\cite{breakj,takayanagi},
a closer comparison between the
predictions of simple quasi one-dimensional theories and experimental
results is now at hand. Refs. \cite{breakj,takayanagi} provide two
different examples of the progress made in the experimental 
achievement of a superconducting quantum point-contact (SQPC).
This possibility has provoked a renewed interest
in more detailed and quantitative analysis of models involving a few
conducting channels. It turns out that, in spite of its apparent simplicity,
the case of a single channel SQPC still 
contains non-trivial physical behavior at certain regimes.
Single mode contacts have been analyzed recently by different 
authors \cite{Shumeiko,gphi,Averin} who have obtained new
quantitative results for the case of small applied voltage where multiple
Andreev reflections (MAR) play a crucial role. 

Some of these recent works \cite{gphi,Averin} illustrate  
the existence of a variety of regimes controlled by the contact 
transmission and quasiparticle damping in the small bias limit. 
A full understanding of the different regimes, together with
the transitions between 
them remains to be explored. This will be thoroughly analyzed in this paper.

Traditionally, quantum transport in microelectronic devices has been 
mainly addressed by two different approaches:
one is based on the scattering picture first introduced
by Landauer \cite{Landauer}
in which the transport properties are expressed in terms of transmission
and reflection scattering amplitudes. This approach is generally used in
a phenomenological way by replacing the device with a simple
scattering structure.
The natural extension of this picture to the superconducting case was
provided by the already mentioned BTK model.

A different point of view arises when the problem is analyzed starting
from a microscopic Hamiltonian.
We shall very generally call ``Hamiltonian approach'' 
the theories which take this starting
point.
The origin of this approach can be traced back to the work by Bardeen
who introduced the tunnel Hamiltonian approximation for describing a
tunnel junction \cite{Bardeen}. Most of the calculations
based on this tunnel Hamiltonian were restricted to the lowest order
transport processes like in the calculation of the Josephson current in
a S-I-S junction \cite{Ambegaokar}. Multi-particle tunneling 
processes (MPT) were first
discussed by Schrieffer and Wilkins \cite{MPT} as a possible explanation
for the observed subgap structure in superconducting tunnel junctions. 
The contributions of these higher order processes were found to be
divergent, which has led to the quite extended belief that the
Hamiltonian approach is pathological except for describing the lowest 
order tunneling processes.

One of the aims of the present work is to show that starting from a very
simple model Hamiltonian describing a single channel contact it
is possible to obtain results for N-S and S-S contacts
in agreement with those provided by
scattering theory. As will be shown below inclusion of
higher order processes up to infinite order eliminates the pathologies
associated with finite order perturbation theory. 

On the other hand, the Hamiltonian approach in combination with 
non-equilibrium Green function techniques presents its own
advantages. In recent publications we have shown how this approach can
be generalized for dealing with situations where self-consistency of
the superconducting order parameter is needed \cite{joslar}.
Moreover, the
formulation in terms of Green functions is specially well suited
for dealing with
correlation effects when strong e-e interactions are present \cite{dot}.

The paper is organized as follows: In section II the general model for a 
superconducting weak link in a local representation is
introduced. In particular, we discuss how this model can be simplified 
in order to
describe a single mode contact. We then introduce the
non-equilibrium Green function formalism by means of which the
total current through the contact can be expressed. 
In section III we analyze the more simple N-N and N-S contacts.
The study of the N-N case allows us to define the contact transmission 
coefficient in terms of the microscopic parameters of our model. 
The complete correspondence with the results of the BTK scattering theory 
is then established by analyzing the N-S contact. For this last case we
derive an analytical expression for the total current as a function of
the contact transmission. 
Section IV is devoted to the S-S case. We first show how the problem of
the calculation of the ac current components can be reduced to the
evaluation of an algebraic set of linear equations. 
This allows us to study in detail the general features of the
dc and ac $I-V$ characteristics
for the whole range of voltages and transmissions. We finally 
concentrate on the limit $V \ll \Delta$ where the
more interesting and less understood physics takes place. 
Section V is devoted to some concluding remarks.

\

\section{Model and method}

In this paper we shall consider  ``point contact-like" geometries consisting
very generally of two wide (3D or 2D) electrodes connected by a narrow 
constriction. The constriction length, $L_C$, is assumed to be much smaller
than the superconducting coherence length and its width, $W_C$, comparable to 
the Fermi wavelength , $\lambda_{F}$. For a sufficiently short constriction 
the detailed form of the self-consistent order parameter and electrostatic
potential in the region between the electrodes becomes irrelevant 
\cite{joslar}
allowing us to represent them by simple step functions. On the other hand, 
the condition $W_C \sim \lambda_{F}$ implies that there is only a reduced
number of quantum transverse channels through the constriction. One can
think of different experimental realizations of such a physical situation,
like for instance the recently developed atomic size break-junctions 
\cite{breakj} and the split-gate S-2DEG-S
SQPC of Takayanagi et al. \cite{takayanagi}. These two situations are
schematically represented in Fig. 1.

The analysis of the transport and electronic properties of such a system 
would involve, for the most general case where both electrodes are
superconductors, the solution of the time-dependent Bogoliubov-de Gennes 
(BdeG) equations \cite{BdeG}. 
In previous works we have developed an approach to 
these kind of problems in which the BdeG equations are formulated
in a site representation \cite{joslar}. This representation can be viewed 
either as a tight-binding description of the electronic states or as a 
discretization of the BdeG equations. The first case would be more
suitable for describing systems like an atomic size contact (Fig. 1 a),
while the second could be used to represent constrictions involving 
a 2DEG in semiconducting heterostructures (Fig. 1 b). In the absence of applied
fields the mean-field Hamiltonian giving rise to the BdeG equations
would take the form 

\begin{equation}
\hat{H} = 
\sum_{i ,\sigma} (\epsilon_{i} - \mu) c^{\dagger}_{i \sigma} c_{i \sigma} + 
\sum_{i \neq j ,\sigma} t_{ij} c^{\dagger}_{i} c_{j} + 
\sum_{i} (\Delta^{*}_{i} c^{\dagger}_{i \downarrow} c^{\dagger}_{i \uparrow} + 
\Delta_{i} c_{i \uparrow} c_{i \downarrow}) ,
\end{equation}

\noindent
where $i,j$ run over the sites used to represent the system. By 
appropriately choosing 
the different parameters $\epsilon_{i}$, $t_{ij}$ and
$\Delta_{i}$ one can model different junction geometries \cite{joslar}.
In order to analyze the case of biased contacts it is convenient to
simplify this model Hamiltonian taking the step-like behavior of
the order parameter and electrostatic potentials in a point contact geometry
into account.
To do this we take the complex order parameter and the chemical potentials
as constants on the left and right electrodes (denoted by 
($\Delta_{L}, \mu_{L}$) and ($\Delta_{R}, \mu_{R}$) respectively).
The case of a single quantum channel connecting both
electrodes can then be described by the following Hamiltonian 

\begin{equation}
\hat{H} = \hat{H}_{L} + \hat{H}_{R} +
 \sum_{\sigma} (t c^{\dagger}_{L \sigma} c_{R \sigma} + 
t^{*} c^{\dagger}_{R \sigma} c_{L \sigma})
- \mu_{L} \hat{N}_{L} - \mu_{R} \hat{N}_{R} ,
\end{equation}

\noindent
where $\hat{H}_{L}$ and $\hat{H}_{R}$ correspond to the uncoupled 
electrodes while the hopping term describes the electron transfer
processes between the outermost sites on both electrodes.
We would like to stress that, although this is a very simple model
Hamiltonian (formally equivalent to a tunnel Hamiltonian), 
it contains the relevant physics of a quantum point 
contact which depends essentially on the contact normal transmission 
coefficient. In our model this transmission can vary between $\sim 0$
(tunnel limit) and $\sim 1$ (ballistic regime) as a function of the coupling 
parameter $t$. This will be discussed in section III.

For the case of a symmetric superconducting contact in the presence 
of an applied bias voltage $eV = \mu_L - \mu_R$ it is convenient to 
perform a gauge transformation \cite{Scalapino} by means of which
Hamiltonian (2) adopts the following time-dependent form

\begin{equation}
\hat{H}(\tau) = \hat{H}_{L} + \hat{H}_{R} +
\sum_{\sigma} \left( t e^{i \phi(\tau)/2} c^{\dagger}_{L \sigma} c_{R \sigma} 
+ t^{*} e^{-i \phi(\tau)/2} c^{\dagger}_{R \sigma} c_{L \sigma} \right),
\end{equation}

\noindent
where the superconducting phase difference $\phi(\tau) = \phi_{0} 
+2eV\tau/ \hbar$ appears as a phase factor in the hopping elements.

\subsection{Current in terms of the non-equilibrium Green functions}

Within the previously introduced model,
the average total current through the contact is given by

\begin{equation}
I(\tau)= \frac{i e}{\hbar}  \sum_{\sigma} (t <c^{\dagger}_{L \sigma}(\tau) 
c_{R \sigma}(\tau)> - \; t^{*} <c^{\dagger}_{R \sigma}(\tau) 
c_{L \sigma}(\tau)>) ,
\end{equation}

\noindent
where, depending on the gauge choice, $t$ can include a time dependent 
phase like in Eq. (3). 
The averaged quantities in Eq. (4) can be expressed in terms
of non-equilibrium Keldysh Green functions $\hat{G}^{+-}$
\cite{Keldysh}.
For the superconducting state it is convenient to introduce the $2
\times 2$
Nambu representation in which $\hat{G}^{+-}_{i,j}$ adopts the form
\cite{joslar}

\begin{equation}
\hat{G}^{+-}_{i,j}(\tau,\tau^{\prime})= i \left(
\begin{array}{cc}
<c^{\dagger}_{j \uparrow}
(\tau^{\prime}) c_{i \uparrow}(\tau)>   &
<c_{j \downarrow}(\tau^{\prime}) c_{i \uparrow}(\tau)>  \\
<c^{\dagger}_{j \uparrow}(\tau^{\prime})
c^{\dagger}_{i \downarrow}(\tau)>  &
<c_{j \downarrow}(\tau^{\prime})
 c^{\dagger}_{i \downarrow}(\tau)>
\end{array}  \right) .
\end{equation}

\noindent
In terms of the $\hat{G}^{+-}$, the current is given by 

\begin{equation}
I(\tau) =\frac{2 e}{\hbar} \left[\hat{t} \hat{G}^{+-}_{RL}(\tau,\tau) -
\hat{t}^{*} \hat{G}^{+-}_{LR}(\tau,\tau) \right]_{11}     ,
\end{equation}

\noindent
where $\hat{t}$ is the Nambu representation of the hopping elements

\begin{equation}
\hat{t} = \left(
\begin{array}{cc}
 t   &   0      \\
  0                    &   -t^{*}
\end{array} \right)  .
\end{equation}

The problem then consists in the determination of these  
non-equilibrium Green functions $\hat{G}^{+-}$.
This can be formally achieved by treating the coupling $\hat{t}$ as a 
perturbation. The unperturbed 
Green functions correspond to 
the uncoupled electrodes in equilibrium. For a symmetric contact,
and neglecting finite band-width effects, the uncoupled retarded and
advanced Green functions 
can be expressed as 

\begin{equation}
\hat{g}^{r,a}_{LL}(\omega)= \hat{g}^{r,a}_{RR}(\omega)= 
\frac{1}{W \sqrt{ \Delta^{2} -(\omega \pm i \eta)^{2} }} \left(
\begin{array}{cc}
-\omega \pm i\eta   &    \Delta   \\
\Delta              &     -\omega \pm i\eta 
\end{array}  \right) ,
\end{equation}

\noindent
where $W$ is an energy scale related to the normal density of states
at the Fermi level by $\rho(\epsilon_F) \sim 1/(\pi W)$
and $\eta$ is a small energy 
relaxation rate that takes into account the damping of the quasi-particle 
states due to inelastic processes inside the electrodes \cite{comment}. 
On the other hand, the 
unperturbed $\hat{g}^{+-}(\omega)$ satisfy the relation 

\begin{equation}
\hat{g}^{+-}(\omega)= 2 \pi i \hat{\rho}(\omega) n_F(\omega) ,
\end{equation}

\noindent
where $\hat{\rho}(\omega)=(1/\pi) Im[\hat{g}^{a}(\omega)] $ 
and $n_F(\omega)$ is the Fermi function.

For the coupled system, the functions $\hat{G}^{+-}_{i,j}$ can be obtained 
from the retarded and advanced functions by means of the integral equation

\begin{eqnarray}
\hat{G}^{+-}(\tau,\tau^{\prime}) & = & \int d \tau_{1} d \tau_{2} \left[ 
\hat{I} \delta(\tau-\tau_1) + \hat{G}^{r}(\tau,\tau_1) 
\hat{\Sigma}^{r}(\tau_1)
\right] \hat{g}^{+-}(\tau_1-\tau_2) \times \nonumber \\
& & \left[
\hat{I} \delta(\tau_2-\tau^{\prime}) + \hat{\Sigma}^{a}(\tau_2)
\hat{G}^{a}(\tau_2,\tau^{\prime}) \right] ,
\end{eqnarray} 

\noindent
where $\hat{G}^{r,a}$ can be derived from their corresponding Dyson equations 

\begin{equation}
\hat{G}^{r,a}(\tau,\tau^{\prime}) = \hat{g}^{r,a}(\tau - \tau^{\prime}) + 
\int d \tau_{1} \hat{g}^{r,a}(\tau - \tau_{1}) \hat{\Sigma}^{r,a}(\tau_1) 
\hat{G}^{r,a}(\tau_{1},\tau^{\prime}).
\end{equation}

\noindent
In the above equations the self-energies take the simple form 
$\hat{\Sigma}^{r,a}_{LL}=\hat{\Sigma}^{r,a}_{RR}=0$ and 
$\hat{\Sigma}^{r,a}_{LR}=(\hat{\Sigma}^{r,a}_{RL})^* = \hat{t}$.

In sections III and IV details of the solution of these integral 
equations are given.

\

\section{N-N and N-S contacts} 

In this section we briefly review the N-N and N-S cases. The analysis of 
the N-N case permits to define the normal transmission coefficient
of the contact in terms of the microscopic parameters of the model, which is 
necessary for making contact with the scattering approach. The 
equivalence of both approaches is illustrated by comparing our results 
for the N-S contact with those of the BTK model \cite{BTK}. 

As we shall deal 
with a  stationary situation it is convenient to adopt the
time-independent formulation based on Hamiltonian (2). In this
case the Green functions depend on the difference of its temporal arguments
and integral equations (10) and (11) become simple algebraic equations when 
Fourier transformed. Then we have

\begin{equation}
\hat{G}^{r,a} (\omega)= \hat{g}^{r,a} (\omega) +
\hat{g}^{r,a} (\omega) \hat{\Sigma}^{r,a}(\omega)  \hat{G}^{r,a} (\omega)
\end{equation}

\begin{equation}
\hat{G}^{(+-),(-+)} (\omega) = \left[ \hat{I} + \hat{G}^{r} (\omega) 
\hat{\Sigma}^{r}(\omega) \right] \hat{g}^{(+-),(-+)} (\omega) 
\left[ \hat{I} + \hat{\Sigma}^{a}(\omega) \hat{G}^{a} (\omega) \right].
\end{equation}

\noindent
where $\hat{g}^{-+}(\omega) =  - 2 \pi i \hat{\rho}(\omega) 
[1 - n_F(\omega) ]$.
On the other hand, it is convenient to rewrite expression (6) 
for the current as \cite{joslar,Caroli}

\begin{equation}
I=\frac{2 e}{h} t^{2} \int^{\infty}_{-\infty} 
\left[ g^{+-}_{LL,11} G^{-+}_{RR,11}(\omega) -
g^{-+}_{LL,11} G^{+-}_{RR,11}(\omega) \right] d \omega ,
\end{equation}

\noindent
where we have assumed that the left electrode is in the normal state, 
while the right one can be either normal or superconducting. 

Let us first analyze the case where both electrodes are in the normal 
state. Details on the calculations of the Green functions for this case 
have been given elsewhere \cite{contact}. As in the scattering approach, the
current can be written as

\begin{equation}
I=\frac{2 e}{h} \int^{\infty}_{-\infty} T(\omega,V)  
\left[ n_F(\omega - eV) - n_F(\omega) \right] d \omega ,
\end{equation}

\noindent
where $T(\omega,V)$ is an energy-dependent transmission coefficient 
which is given by

\begin{equation}
T(\omega,V)=\frac {4 \pi^{2} t^{2} \rho_{LL} (\omega -eV) \rho_{RR} (\omega)} 
{|1 - t^{2} g_{LL} (\omega -eV) g_{RR} (\omega)|^{2}} .
\end{equation}

This expression can be further simplified assuming that the normal 
state spectral densities are constant over an energy scale $W \gg V,
\Delta$. This is consistent with the assumptions leading to Eq. (8).
Within this energy scale the coefficient 
$T(\omega,V)$ becomes an energy-independent quantity $\alpha$

\begin{equation}
T(\omega,V) \simeq \frac{4 t^{2}/ W^{2}} {(1 + t^{2}/W^{2})^{2}} 
\equiv \alpha .
\end{equation}

The normal linear conductance of this single mode contact
is then given by the Landauer formula 
$G_{NN} = (2e^{2}/h) \alpha$. Notice that $\alpha$ 
can vary between zero and one as a function of $t$. 
The $\alpha \rightarrow 0$ limit is reached both for $t/W \ll 1$ and
for $t/W \gg 1$, while 
the ballistic limit, i.e. $\alpha \sim 1$, is reached when 
$t/W \sim 1$. 

It is worth to further clarify the role of the hybridization parameter 
$t$ within our point contact model. Although in the context of the
tunnel Hamiltonian approach it has been customary to
identify $\sim t^2$ with the transmission
probability, this strictly holds for the tunnel regime 
(lowest order perturbation theory in $t$). However, the actual
expression for the contact transparency (Eq. (17)) 
including all order processes is a non-linear
function of $t^2$. While the ballistic condition is achieved for 
$t^2/W^2 \sim 1$, the way in which $\alpha$ approaches unity is completely 
different from $\sim t^2/W^2$, as can be seen from Eq. (17).

In the N-S case ($\Delta_{L} = 0$, $\Delta_{R} = \Delta$) starting from
Eq. (12) and (13) after some simple algebra
we obtain the 
following expression for the current as the sum of four 
different contributions $I = I_{1} + I_{2} + I_{3} + I_{A}$, where 

\begin{eqnarray}
I_{1} & = & \frac{8 e}{h} \pi^{2} t^{2} \int^{\infty}_{-\infty} 
d \omega |1+ t G^{r}_{RL,11}(\omega)|^{2} 
\rho_{LL,11}(\omega-eV) \rho_{RR,11}(\omega) \left[ 
n_F(\omega-eV)-n_F(\omega) \right] \nonumber \\
I_{2} & = & - \frac{16 e}{h} \pi^{2} t^{2} \int^{\infty}_{-\infty} 
d \omega Re \left\{ t G^{a}_{LR,21}(\omega) \left[
1 + t G^{r}_{RL,11}(\omega) \right] \right\} \times \nonumber \\
& & \hspace{2cm} \rho_{LL,11}(\omega-eV) \rho_{RR,12}(\omega) \left[ 
n_F(\omega-eV)-n_F(\omega) \right] \nonumber \\
I_{3} & = & \frac{8 e}{h} \pi^{2} t^{4} \int^{\infty}_{-\infty} 
d \omega |G_{RL,12}(\omega)|^{2}
\rho_{LL,11}(\omega-eV) \rho_{RR,22}(\omega) 
\left[ n_F(\omega-eV)-n_F(\omega) \right] \nonumber \\
I_{A} & = & \frac{8 e}{h} \pi^{2} t^{4} \int^{\infty}_{-\infty} 
d \omega |G_{RR,12}(\omega)|^{2}
\rho_{LL,11}(\omega-eV) \rho_{LL,22}(\omega+eV)
\left[ n_F(\omega-eV) - n_F(\omega+eV) \right] . \nonumber \\
& &
\end{eqnarray}

Written in this form, each contribution has a clear interpretation in 
terms of elementary processes that can be identified by inspection of
the intervening spectral densities. Although 
there is a formal resemblance of the
above expressions with those of tunnel theory \cite{Mahan} Eq. (18)
contains all possible processes up to infinite order in $t$. 
Thus, $I_{1}$ corresponds to
normal electron transfer between the electrodes, while $I_{2}$ corresponds 
also to a net transfer of a single electron with creation or annihilation of
pairs as an intermediate state. 
On the other hand, $I_{3}$ arises from processes where an electron in 
the normal electrode is converted into a hole in the superconducting
side, i.e. processes with ``branch crossing" in the BTK language. 
Finally, $I_{A}$ arises from Andreev reflection processes in which
an electron (with an associated spectral weight 
$\rho_{LL,11}(\omega-eV)$) is transmited from the left to the right 
electrode with a hole reflecting backwards into the normal electrode
(with an associated spectral weight $\rho_{LL,22}(\omega+eV)$)
while a Cooper pair is created in the superconducting side with a
probability proportional to $|G_{RR,12}(\omega)|^{2} $.

 As expected, the only non-zero contribution for $eV< \Delta$   
is $I_{A}$, while all processes contribute for $eV>\Delta$. With the
same simplifying assumptions leading to Eq. (17) the differential 
conductance at zero temperature adopts the simple form

\begin{eqnarray}
G_{NS}(V) & = & \frac{4 e^{2}}{h}  
\frac{\alpha^{2}}{(2- \alpha)^{2} - 4(1- \alpha)
(\frac{eV}{\Delta})^{2}} ; \hspace{1cm} eV \leq \Delta  \nonumber \\
G_{NS}(V) & = & \frac{4 e^{2}}{h} 
\frac{\alpha}{\alpha + (2- \alpha) \sqrt{1-(\frac{\Delta}{eV})^{2}}} 
; \hspace{1cm} eV > \Delta .
\end{eqnarray}

This expression can be shown to be equivalent to the one obtained from 
the BTK model \cite{BTK} with the correspondence $Z = [1 -
(t/W)^2]/(2t/W)$, where $Z$ is the dimensionless phenomenological
parameter controlling the barrier height in the BTK model.

It is worth noticing that the differential conductance rises from 
$(4 e^2/h) \alpha^2/(2 - \alpha)^2$ at $V = 0$ to the value $4 e^2 /h$
at $eV = \Delta$, this last value being independent of the contact
transmission.
This result
can never be obtained within any finite order perturbative approximation 
in $t$, but requires an infinite order calculation \cite{comment2}. 
This sort of non-perturbative features are also very important in the 
S-S case \cite{Hasselberg}, as will be discussed in the next section.

Finally, it is possible to obtain from Eq. (18)
an analytical expression of 
the total current at zero temperature. For $eV > \Delta$
this can be written as $I = I_1 + I_2$ with
 
\begin{eqnarray}
I_{1} &=& \frac{e \Delta}{h} \frac{\alpha^{2}}{(2-\alpha) \sqrt{1-\alpha}}
\ln \left[ \frac {1 + \left[ \frac{2 \sqrt{1-\alpha}}{2-\alpha} \right] }
{1 - \left[ \frac{2 \sqrt{1-\alpha}}{2-\alpha} \right] } \right]
\nonumber \\
I_{2}(x) &= & \frac{4e}{h} \Delta \left[ \frac{\alpha^{2}}{4(1-\alpha)} + 
\frac{\alpha}{x \left[ \alpha+(2-\alpha)\sqrt{1- x^2} \right]} - 
\frac{\alpha (2-\alpha)^{2}x}{4(1-\alpha) \left[ \alpha+(2-\alpha) 
\sqrt{1 - x^2} \right]} \right. \nonumber \\
&& + \left.\frac{\alpha^{2}(2-\alpha)}{8(1-\alpha)^{3/2}} \ln \left[ 
\frac{ \left( \frac{x}{1+\sqrt{1- x^2}} + \frac{1}{\sqrt{1-\alpha}} \right) 
\left( 1 - \frac{1}{\sqrt{1-\alpha}} \right)}
{\left( \frac{x}{1+\sqrt{1- x^2}} - \frac{1}{\sqrt{1-\alpha}} \right)
\left( 1 + \frac{1}{\sqrt{1-\alpha}} \right)} \right] \right] ,
\end{eqnarray}

\noindent
where $I_1$ is the gap contribution to the total current, $I_2$ is the 
contribution coming from energies outside the gap and $x=\Delta/eV$. 
This allows to analyze with 
detail the ``excess" current, defined as
$I_{exc}=  \lim_{V \rightarrow \infty} (I_{NS}-I_{NN})$, as a function
of the contact transparency. We find $I_{exc}= I_{exc_{1}} + I_{exc_{2}}$

\begin{eqnarray}
I_{exc_{1}} & = & 
\frac{e \Delta}{h} \frac{ \alpha^{2}}{(2- \alpha) \sqrt{1- \alpha}} 
\ln \left[ \frac{1+ \left[ \frac{2 \sqrt{1- \alpha}}{2- \alpha} \right ]} 
{1- \left[ \frac{2 \sqrt{1- \alpha}}{2- \alpha} \right ]} \right]
\nonumber \\
I_{exc_{2}} & = & \frac{e \Delta}{h} \alpha^{2} \left[ \frac{1}{1- \alpha} + 
\frac{2- \alpha}{2(1- \alpha)^{3/2}} \ln \left[ \frac{1- \sqrt{1- \alpha}} 
{1+ \sqrt{1- \alpha}} \right] \right], \nonumber
\end{eqnarray}
             
\noindent
where $I_{exc_{1}}$ and $I_{exc_{2}}$ are respectively the contributions
coming from energies inside and outside the gap.
As can be easily checked, $I_{exc_{1}}>0$ while $I_{exc_{2}}<0$ the
total excess current being always positive. 
For $\alpha=1$ the well known result \cite{Zaitsev,KBT} for the ballistic
contact $I_{exc}=(8/3) e \Delta/h$ is recovered.

\section{The S-S contact}

\subsection{An efficient algorithm for evaluating the ac current} 

As commented in section II, for the case of a voltage-biased S-S contact
it is convenient to start from
Hamiltonian (3) in which the applied bias is taken into account through
a time-dependent phase factor 
in the hopping element, which in a Nambu representation has the form

\begin{equation}
\hat{t} = \left(
\begin{array}{cc}
 t e^{i \phi(\tau)/2}  &   0      \\
  0                    &   -t e^{-i \phi(\tau)/2}
\end{array} \right)  ,
\end{equation}

\noindent
where $\phi= \phi_{0} + 2eV \tau /\hbar$ is the time-dependent 
superconducting phase difference. 

This explicit time dependence indicates that all dynamic quantities
can be expanded as Fourier series in all possible harmonics of the 
fundamental frequency $\omega_{0}=2eV/ \hbar$ \cite{gphi}.
For instance, the total current can be written as 

\begin{equation}
I(\tau) = \sum_{m} I_{m} e^{im \omega_{0} \tau}.
\end{equation}

We shall now show how these Fourier coefficients, $I_{m}$, can be efficiently
evaluated within the non-equilibrium Green function formalism. Let us first
notice that the non-equilibrium Green functions appearing in Eqs. (10)
and (11)
do not depend only on the difference of their temporal arguments, 
and have therefore a generalized 
Fourier expansion of the form \cite{Arnold,photons}

\begin{equation}
\hat{G}(\tau,\tau^{\prime})= \frac{1}{2 \pi} \sum_{n} \int 
d \omega e^{-i \omega \tau} e^{i (\omega+n \omega_{0}/2) \tau^{\prime}} 
\hat{G}(\omega, \omega+n \omega_{0}/2).
\end{equation}

Hereafter we shall use the notation $\hat{G}_{nm}(\omega)=\hat{G}(\omega+n
\omega_{0}/2,\omega+m\omega_{0}/2)$.
Different Fourier components $\hat{G}_{nm}$ are related 
by $\hat{G}_{nm}(\omega)=\hat{G}_{n-m,0}(\omega+m\omega_{0}/2)$.
For the particular gauge choice adopted here, it is useful
to express all quantities in terms of a renormalized hopping which satisfies 
its own Dyson equation

\begin{equation}
\hat{T}^{a,r}(\tau,\tau^{\prime})= \hat{t}(\tau) \delta(\tau-\tau^{\prime}) + 
\int d \tau_{1} d \tau_{2} \hat{t}(\tau) \hat{g}^{a,r}(\tau-\tau_{1}) 
\hat{t}^{\dagger} (\tau_{1}) \hat{g}^{a,r}(\tau_{1}-\tau_{2})
\hat{T}^{a,r}(\tau_{2},\tau^{\prime}).
\end{equation}

This quantity can be viewed as the total hopping amplitude arising from 
summing up all processes in which one electron is transferred. Clearly,
it is formally equivalent to use a renormalized hopping 
instead of renormalized propagators 
as they are linked by relations like
$\hat{G}_{LL} (\tau, \tau') \hat{t}(\tau') = \int d \tau_1
\hat{g}_{L}(\tau - \tau_1) \hat{T}_{LR} (\tau_1, \tau')$.
The current components can now be expressed in terms of the 
renormalized hopping elements $\hat{T}^{a,r}_{nm} (\omega)$ as 

\begin{eqnarray}
I_m &= & \frac{2 e}{h} \int d\omega \sum_n
\left[ \hat{T}_{0n}^r \hat{g}_{nn}^{+-}
\hat{T}_{nm}^{r\dagger} \hat{g}_{mm}^a
- \hat{g}_{00}^{r} \hat{T}_{0n}^r
\hat{g}_{nn}^{+-} \hat{T}_{nm}^{r \dagger}
\nonumber \right. \\
&& \left. + \hat{g}_{00}^{r} \hat{T}_{0n}^{a \dagger}
\hat{g}_{nn}^{+-} \hat{T}_{nm}^{a}
 - \hat{T}_{0n}^{a \dagger} \hat{g}_{nn}^{+-}
\hat{T}_{nm}^{a} \hat{g}_{mm}^{a} \right]_{11},
\end{eqnarray}
where we have eliminated the site indexes $L$ and $R$ 
in the uncoupled Green functions
due to the left-right symmetry of the contact. 

The problem is then reduced to that of the evaluation of the components 
$\hat{T}_{nm}$. From Eq. (24) it can be shown 
that the components $\hat{T}_{nm}$ (both retarded and advanced parts)
satisfy a set
of linear equations of the form 

\begin{equation}
\hat{T}_{nm} =\hat{t}_{nm} + \hat{\epsilon}_{n} \hat{T}_{nm} +
\hat{V}_{n,n-2} \hat{T}_{n-2,m} + \hat{V}_{n,n+2} \hat{T}_{n+2,m}.
\end{equation}

These equations are mathematically equivalent to those describing the motion
of electrons in a tight-binding linear chain with ``site energies" ,
$\hat{\epsilon}_{n}$, and ``nearest-neighbor couplings",
$\hat{V}_{n,n-2}$ and $\hat{V}_{n,n+2}$.
The detailed expression of $\hat{\epsilon}_{n}$ and $\hat{V}_{n,m}$ in 
terms of the unperturbed Green functions
are given in Appendix A. This analogy allows us to obtain the Fourier
coefficients $\hat{T}_{nm}$
using standard recursion techniques. One can show (see Appendix 
A) that the following recursive relation holds

\begin{equation}
\left\{ \begin{array}{lr}
\hat{T}_{n+2 ,m}(\omega) = \hat{z}^{+} \left[\omega+(n-1)\omega_{0}
\right] \hat{T}_{n m}(\omega) & ,n\geq 1 \\
\hat{T}_{n-2 ,m}(\omega) = \hat{z}^{-} \left[ \omega+(n+1) \omega_{0} 
\right] \hat{T}_{n m}(\omega) & ,n \leq -1
\end{array} \right.
\end{equation}

\noindent
where the transfer matrix $\hat{z}^{\pm}$ satisfy the equation 

\begin{equation}
\hat{z}^{\pm}(\omega)= \left[ \hat{I} - \hat{\epsilon}_{\pm 3} - 
\hat{V}_{\pm 3, \pm 5} \hat{z}^{\pm}(\omega \pm \omega_{0}) \right]^{-1}.
\end{equation}

Clearly, as the transfer matrix $\hat{z}^{\pm}$ connects
consecutive harmonics 
of $\hat{T}$, it can be viewed as a generating function which introduces 
the effect of a unitary Andreev reflection process.
The problem has been reduced to the calculation of only two matrix 
coefficients like, for instance, $\hat{T}_{1,0}$ and $\hat{T}_{-1,0}$ 
as a starting point for the generating Eqs. (27) (see Appendix A for details). 

In summary, the basic mathematical difficulty lies in the evaluation of the 
transfer matrix functions $\hat{z}^{\pm}$ from Eq. (28). 
Although Eq. (28) looks simple, it
is nevertheless hard to solve analytically for arbitrary values of $V$.
The analytical results presented in this paper are 
limited to the $eV/\Delta \rightarrow 0$ and $eV/\Delta \rightarrow \infty$ 
cases where some simplifying relations hold. 
For intermediate voltages, an accurate numerical solution of Eq. (28) 
can be obtained.

\subsection{Analysis of ac and dc $I-V$ characteristics}

In this subsection we analyze the general features of the $I-V$ characteristics 
of a S-S contact obtained using our formalism. Let us
start by briefly discussing the dc current, $I_{0}(V)$, for different values 
of the transmission, as shown in Fig. 2. Although the overall qualitative 
features of these curves have been known since the works of Octavio et al.
\cite{OBTK}, Zaitsev \cite{Zaitsev} and Arnold \cite{Arnold}, more 
quantitative and detailed analysis are being reported in recent publications
\cite{Shumeiko,Averin}. The results of Fig. 2 are in agreement
with those reported recently by Averin and Bardas \cite{Averin} which were
obtained using the scattering approach.

Two relevant features of these curves are the subharmonic gap structure for
$eV<2 \Delta$ and the excess current for $eV \gg \Delta$.
As can be observed, the subgap structure becomes
progressively more pronounced with decreasing transmission. Eventually, when
$\alpha \ll 1$ the current steps at positions 
$eV \sim 2 \Delta /n$ can be clearly resolved. 
In this limit one can isolate the 
elementary processes which give rise to these steps. It can be shown 
that in the tunnel limit the $n$-th step can be calculated as

\begin{equation}
\delta I^{(n)}_0  =  \lim_{eV \rightarrow 2 \Delta^+/n}
\frac{8e}{h} \pi^2 n t^{2n} \int^{neV-\Delta}_{\Delta} 
d \omega \left[ \prod^{n-1}_{i=1} |g_{12}(\omega-ieV)|^2 \right] 
\rho_{11}(\omega) \rho_{22}(\omega-neV) . 
\end{equation}

By comparison of Eq. (29) with the expression of $I_A$ in Eq. (18)
it becomes clear that 
the steps inside the gap are due to the opening of a 
new Andreev reflection 
channel whenever $eV=2\Delta/n$. Calculation of the integrals in Eq. (29) 
leads to

\begin{equation}
\delta I^{(n)}_0= \frac{e\Delta \alpha^n}{\hbar} 
\left( \frac{2n}{4^{2n-1}} \right) \left( \frac{n^n}{n!} \right)^2 ,
\end{equation}

\noindent
in agreement with the recent prediction of Bratus et al. \cite{Shumeiko}. 
On the other hand, when $\alpha \rightarrow 1$ 
the subgap structure is completely washed out and there
appears an excess current even in the small bias limit.

The other relevant feature of the dc current, namely the 
large voltage excess current, can
be analytically evaluated within our model for any transmission value. The
main simplification in this limit comes from the fact that only the lowest 
order Andreev reflection process gives a significant contribution to the
excess current \cite{OBTK}. This implies that one can truncate the system 
of equations (26) for harmonic indexes $n>1$, the resulting simplified system
can then be solved explicitly for $\hat{T}_{1,0}(\omega)$ and 
$\hat{T}_{-1,0}(\omega)$ (for details see Appendix B). As shown in this 
Appendix the simple result $I^{SS}_{exc}=2 I^{NS}_{exc}$ is obtained for any
value of the transmission. Although this is the expected physically sound
result, to our knowledge, it has not been shown explicitly before
except for the ballistic case \cite{Zaitsev,KBT}. Moreover, some 
authors have reported the existence of a negative excess current for low
transmissions \cite{Flens.} which seems to be in  contradiction with the 
above result. However, one should notice that the excess current as defined
above is an asymptotic quantity (only valid in the $eV/ \Delta \rightarrow 
\infty$ limit). When corrections of order $\Delta /eV$ are taken 
into account one actually can have a defect instead of an excess current
for sufficiently low transmission.

The algorithm described in subsection A allows an efficient evaluation
of the higher order ac components of the current. 
For the following analysis we decompose the ac current, Eq. (22), into 
its dissipative and non-dissipative contributions given respectively by

\begin{equation}
I_D= I_0 + \sum_m I^{D}_m \cos(m \omega_0 \tau)
\end{equation}

\noindent
and

\begin{equation}
I_S= \sum_m I^{S}_m \sin(m \omega_0 \tau) ,
\end{equation}

\noindent
where $I^{D}_m= 2 Re(I_m)$ and $I^{S}_m=-2 Im(I_m)$.

The results obtained for the first three $I^{D}_m$ and $I^{S}_m$ 
components are depicted in Fig. 3 and Fig. 4.
As can be observed, these components become exponentially small
for bias voltages
larger than $\Delta/n$. On the contrary, when $eV < \Delta/n$ the
decay of the ac components with increasing $n$ becomes slower.
The analysis of the higher order components reveals a decay 
for $eV < \Delta/n$ close to a inverse power law.
As a consequence of this slow decay, one is forced to take 
an increasing number of ac components into account in order
to adequately describe the
behavior at small bias. This will be the subject of the next
subsection.

\subsection{Small bias regime}

In this subsection we concentrate on the $eV/ \Delta \rightarrow 0$ case,
which turns out to exhibit a remarkable variety of different regimes 
according to the values of the parameters $\alpha \Delta$ and the
inelastic scattering rate $\eta$.
As it is well known, the main difficulty for 
obtaining quantitative results in this limit lies in the fact that the 
number of MAR contributing to the current grows with decreasing $V$ as 
$\sim \Delta/eV$ \cite{Arnold}. Furthermore, the amplitudes of these 
multiple processes 
do not decay when $(V,\eta) \rightarrow 0$ leading to the
appearance of divergencies in the  
perturbative expansion in the coupling $\hat{t}$
\cite{gphi}. Again, a complete summation of the perturbative 
series is needed in order to regularize these divergencies. 
An additional difficulty arises, as will be discussed below,
from the fact that the limits   
$V \rightarrow 0$ and $\eta \rightarrow 0$ do not actually commute. 

When decreasing the bias voltage two main situations can be reached 
depending on the strength of the inelastic scattering rate at the 
superconducting electrodes: 
the case of $eV/\Delta \rightarrow 0$ and finite $\eta$ has been
discussed by the present authors in recent publications \cite{gphi};
on the other hand, the case of small $V/\Delta$ and negligible 
$\eta$ has been recently addressed to by Averin and Bardas 
\cite{Averin}.
In what follows we summarize the main results for both regimes and analyze 
the conditions for their actual observability in a real SQPC.

Within our formalism, analytical results in the small bias
limit become feasible since the transfer matrix $\hat{z}^{\pm}(\omega)$ 
tends to a scalar quantity having the form of a simple phase factor inside the 
gap region. As shown in Appendix C for $(\eta, V) \rightarrow 0$ we find 

\begin{equation}
\hat{z}^{\pm}(\omega) = 
z(\omega)= e^{i \varphi(\omega)} \; \; ,  \hspace{1cm}
\Delta \sqrt{1-\alpha} \leq |\omega| \leq \Delta ,
\end{equation}

\noindent
where 

\begin{eqnarray}
\varphi(\omega)& = & \arcsin \left( \frac{2}{\alpha \Delta^2} 
\sqrt{\Delta^2 - \omega^2} \sqrt{\omega^2 - (1-\alpha) \Delta^2} \right).
\nonumber
\end{eqnarray}

As the multiple Andreev processes are generated by successive applications of 
$\hat{z}(\omega)$, Eq. (33) indicates that these processes do not decay in this
limit inside the gap region. This infinite series of MAR gives rise to the 
well known bound states spectrum of a current-carrying SQPC at zero bias 
voltage \cite{Furusaki}. As shown in Appendix C, the positions of these 
bound states are
determined by the condition $\varphi(\omega)=\phi$. The presence of a small 
but finite $\eta$ or $V$ (whichever is larger) introduces an effective 
damping into the otherwise infinite series of MAR.

When this effective damping is due to a finite $\eta$ ($eV \ll \eta$) a linear 
regime can be defined where the total current is given by $I_S(\phi) + 
G(\phi)V$, $G(\phi)$ being a phase-dependent linear conductance
\cite{Zaitsev,gphi}. Within 
this linear regime, one can identify two different sub-regimes according 
to the ratio $\eta/ \alpha \Delta$. The case $\eta/ \alpha \Delta \ll 1$ 
corresponds to a situation where MAR are very weakly damped and give
the dominant contribution to the current. The physical picture one can 
have in mind is that of electrons and holes Andreev-reflecting between the 
electrodes for a very long time before being inelastically scattered 
\cite{Zaikin}. In order to illustrate the dominant 
contribution of the processes inside the gap for the weakly damped case,
we represent in Fig. 5 the 
current density corresponding to
the $I_0$ component for three values of the
transmission. Three important features of this linear regime are
displayed in this figure: firstly, the current density inside the gap
increases as $\sim 1/\eta$ therefore giving the dominant contribution 
in the weakly damped case; secondly, there is a region inside
the gap of width $2 \Delta \sqrt{1 - \alpha}$ in which the current density
vanishes. This is the forbidden energy region for bound states at a
given transmission. Finally, in Fig. 5 (c) one can observe that the
contribution of the continuum outside the gap becomes important as
$\alpha < \eta/\Delta$.      
In the limit $\eta/ \alpha \Delta \gg 1$ a second sub-regime is reached
where the contributions of MAR are heavily damped and the current is 
dominated by single quasi-particle tunneling processes. The transition 
between these two sub-regimes has been analyzed with detail in Refs.
\cite{gphi}.

In order to identify the actual sub-regime for a real SQPC an estimation
of the order of magnitude of $\eta$ is needed. 
In Ref. \cite{comment} 
$\eta$ is estimated from the electron-phonon interaction
to be a small fraction of the gap for 
traditional superconductors.
Thus, our theory predicts that a real SQPC would  
generally fall into the weakly damped case except for extremely low
transmissions. 

For this sub-regime the supercurrent $I_S$ and the linear conductance
$G(\phi)$ can be obtained analytically as discussed in Appendix C.
In particular, for $G(\phi)$ one obtains  

\begin{equation}
G(\phi) = \frac{2e^2}{h} \frac{\pi}{16 \eta} \left[ \frac{\Delta \alpha
 \sin \phi}{\sqrt{1-\alpha \sin^2 (\phi/2)} } \mbox{sech}
(\frac{\beta \omega_{S}} {2}) \right]^2 \beta , 
\end{equation}

\noindent
where $\omega_S$ is the position of the bound states inside the gap and
$\beta = 1/k_B T$. This expression for large transmission and small 
temperatures gives a phase-dependence which is in qualitative agreement 
with the few available experimental results, performed in
non-mesoscopic contacts \cite{Rifkin}. The unusual phase-dependence of
Eq. (34) which deviates strongly from the $\cos(\phi)$ form predicted by
the standard tunnel theory may explain the old controversy
between tunnel theory and experiments known as the $\cos(\phi)$ problem 
\cite{Zorin,Barone}.
   
On the other hand, when the truncation of the infinite series of 
MAR is caused by a finite $V$ (with negligible $\eta$), analytical 
results have only been obtained in the quasi-ballistic limit, i.e. 
$\alpha \rightarrow 1$. A closer inspection of the $I-V$ curves in the
small bias region and for $\alpha \sim 1$ reveals that the
supercurrent components decay exponentially from its value at $V = 0$
with a collapsing width $\sim (1 - \alpha) \Delta$. 
This is illustrated in Fig. 6 (a) where a blow up of the behavior of
$I^S_1$ for small bias voltages and quasi-ballistic transmissions is 
shown. 
In the limit $\alpha \rightarrow 1$ the supercurrent 
becomes a delta function at $V = 0$. 
On the contrary, the dissipative components in this same limit tend
to a finite value outside the region of width $\sim (1 - \alpha) \Delta$.
This behavior is shown in Fig. 6 (b) where $I^D_1$ is plotted in the
same magnified scale as $I^S_1$. 
The summation of these dissipative components for $\alpha = 1$ 
and very small $V$ yields (see Appendix C)

\begin{equation}
I_D(\phi)= \frac{e \Delta}{\hbar}|\sin(\phi/2)| \mbox{sign} V ,
\end{equation}

\noindent
in agreement with the result recently derived by Averin and Bardas
\cite{Averin}.  
The existence of a region of decreasing width $V \sim (1 - \alpha) \Delta$ 
in which this crossover from supercurrent to dissipative current takes
place can be associated with the collapse of the
forbidden region for MAR inside the superconducting gap taking place
when $\alpha \rightarrow 1$. In this way, when $V$ is small compared to
the width of the forbidden region 
the excitation of quasi-particles from states at $\omega < -\Delta 
\sqrt{1 - \alpha}$ into states at $\omega > \Delta \sqrt{1 -
\alpha}$ is negligible and there is no appreciable dissipative
current; whereas the opposite situation holds for $V > \sqrt{1 - \alpha}
\Delta$. Averin and Bardas \cite{Averin} have described this crossover
as a Landau-Zener transition in which the non-dissipative and the
dissipative components scale with $\alpha$ and $V$ 
as $(1 - p)$ and $p$ respectively, where $p = \exp{[-\pi (1 - \alpha)
\Delta/eV]}$. The numerical results for sufficiently small $eV/\Delta$ and 
$(1 - \alpha)$ are well fitted by these scaling laws. However, a careful
analysis reveals that their range of validity around $V=0$ and
$\alpha=1$ decreases strongly when increasing the component number.   

In summary, in this small bias limit one can identify four different
sub-regimes depending on the relative values of parameters $\eta$,
$\alpha \Delta$ and $eV$. The prediction of the actual behavior of a
real SQPC in this limit would therefore require a careful estimation of
all these parameters. In this respect, one should keep in mind that 
while $eV$ and $\alpha$ can be varied experimentally in a rather
controlled way, the inelastic scattering rate $\eta$ is an intrinsic property
of the superconducting electrodes much more difficult to control.
The unavoidable presence of some degree of inelastic scattering 
can prevent the actual 
observability of the crossover from non-dissipative to dissipative
behavior described after Eq. (35).
The requirement of $eV \sim (1-\alpha) \Delta$ together with that of
$\alpha \sim 1$ can actually imply $eV < \eta$ which would rather 
correspond to the linear regime.
Finally, when considering the experimental test of all these theoretical
predictions the relevance of noise in a real SQPC should be taken into
account. Recent theoretical predictions \cite{fluct,Ouboter} suggest
that the magnitude of thermal noise in a single mode superconducting
device can be extremely large near ballistic conditions.

\section{Concluding remarks}

In the present work we have presented a Hamiltonian approach for
describing the
transport properties of single mode N-S and S-S contacts.

It has been explicitly demonstrated that this approach is, with some
simplifying assumptions, equivalent to the
phenomenological scattering approach.  
We believe that the present work can help  
clarifying the somewhat recurrent discussion about the
unsuitability of a Hamiltonian approach for obtaining the transport 
properties of a N-S or S-S contact: when performing the calculations up
to infinite order in the coupling $t$ all the unphysical divergences are
eliminated and the results become equivalent to those of scattering theory.

On the other hand, the present approach has been applied to discuss in
detail the dc and ac $I-V$ characteristics of a SQPC. In particular, we
have concentrated on the less understood small bias voltage limit where
one can identify four different sub-regimes depending on the values of
the contact transmission and inelastic scattering rate. Finally, we
have discussed the conditions for the experimental observability of the
theoretical predictions in this limit.    

Although in the present work we have restricted the discussion to the
simplest single-mode case with an energy-independent transmission
coefficient, the general model introduced in section II can describe
more complex situations which may be relevant for a closer comparison 
with recent experiments \cite{takayanagi}. Work along this line is under
progress. 

\acknowledgements
Support by Spanish CICYT (Contract No. PB93-0260) is 
acknowledged. The authors would like to thank A. Lopez D\'avalos for
his helpful remarks on this work.

\appendix
\section{} 

In this Appendix we describe the algorithm for evaluating 
the ac current components in a S-S biased contact.
As it has been pointed out in section IV, we can express the current in terms
of the retarded and advanced
Fourier components of the renormalized hopping $\hat{T}_{nm}(\omega)$
satisfying Eq. (26)

\begin{equation}
\hat{T}_{nm} =\hat{t}_{nm} + \hat{\epsilon}_{n} \hat{T}_{nm} +
\hat{V}_{n,n-2} \hat{T}_{n-2,m} + \hat{V}_{n,n+2} \hat{T}_{n+2,m} ,
\end{equation}

\noindent 
where the matrix coefficients $\hat{\epsilon}_n$ and $\hat{V}_{nm}$ can
be expressed in terms of the uncoupled Green functions as 

\begin{equation}
\hat{\epsilon}_{n} = t^{2} \left(
\begin{array}{cc}
 g_{n+1}g_{n}  &   g_{n+1}f_{n}      \\
 g_{n-1}f_{n}  &   g_{n-1}g_{n}
\end{array} \right)  
\end{equation}

\begin{equation}
\hat{V}_{n,n+2} = -t^{2} f_{n+1} \left(
\begin{array}{cc}
 f_{n+2}  &   g_{n+2}      \\
     0    &   0
\end{array} \right)  
\end{equation}

\begin{equation}
\hat{V}_{n,n-2} = -t^{2} f_{n-1} \left(
\begin{array}{cc}
      0      &   0      \\
   g_{n-2}   &  f_{n-2} 
\end{array} \right)  ,
\end{equation}
                           
\noindent
where $f(\omega) \equiv g_{12}(\omega)=g_{21}(\omega)$ 
and $g(\omega) \equiv g_{11}(\omega)= 
g_{22}(\omega)$ due to electron-hole symmetry.
In the above equations the short-hand notation 
$g_{n}=g(\omega+ n\omega_{0}/2)$ is used.
Moreover, the site indexes in the Green functions have been omitted
since we are considering a symmetric contact.

As commented in section IV, the linear Eqs. (A1) are analogous to those 
describing a tight-binding chain with nearest-neighbor hopping
parameters $\hat{V}_{n,n+2}$ and $\hat{V}_{n,n-2}$.
A solution can then be obtained by
standard recursive techniques. 
It is straightforward to show that the following recursive relations
between the coefficients $\hat{T}_{nm}$ hold

\begin{eqnarray}
\hat{T}_{n+2 ,m}(\omega) & = & \hat{z}^{+}[\omega+(n-1)\omega_{0}] \;
\hat{T}_{n m}(\omega) , \hspace{1cm} n\geq 1 \nonumber\\
\hat{T}_{n-2 ,m}(\omega) & = & \hat{z}^{-}[\omega+(n+1) \omega_{0}] \; 
\hat{T}_{n m}(\omega), \hspace{1cm} n \leq -1
\end{eqnarray}

\noindent
where the transfer matrix $\hat{z}^{\pm}(\omega)$ satisfy the equation 

\begin{equation}
\hat{z}^{\pm}(\omega)= \left[ \hat{I} - \hat{\epsilon}_{\pm 3} -
\hat{V}_{\pm 3, \pm 5} \hat{z}^{\pm}(\omega \pm \omega_{0}) \right]^{-1}.
\end{equation}

\noindent
One can see from Eq. (A6) that $\hat{z}^{+}(\omega)$ and $\hat{z}^{-}(\omega)$ 
are related by $\hat{z}^{-}(\omega,V)= \hat{\sigma}_{x} \hat{z}^{-}(\omega,-V)
\hat{\sigma}_{x}$, where $\hat{\sigma}_{x}$ is the corresponding Pauli matrix.

By virtue of the relation
$\hat{T}_{nm}(\omega)=\hat{T}_{n-m,0}(\omega+m\omega_{0}/2)$,
one can write the current components given by Eq. (25) in terms of
$\hat{T}_{n0}(\omega) \equiv \hat{T}_{n}$.
Using recursive relations (A5) the calculation can be reduced
to a closed system for coefficients $\hat{T}_1$ and $\hat{T}_{-1}$

\begin{eqnarray}
\left[ \hat{I} - \hat{\epsilon}_{1} - \hat{V}_{13} 
\hat{z}^{+}(\omega) \right] 
\hat{T}_{1} &=& \hat{t}_{10} + \hat{V}_{1,-1} \hat{T}_{-1} \nonumber \\ 
\left[ \hat{I} - \hat{\epsilon}_{-1} - \hat{V}_{-1,-3} 
\hat{z}^{-}(\omega) \right] 
\hat{T}_{-1} & = & \hat{t}_{-10} + \hat{V}_{-1,1} \hat{T}_{1}.
\end{eqnarray}

The remaining task is the calculation of the transfer matrix 
$\hat{z}^{+}(\omega)$. It can 
be shown that the solution of Eq. (A6) is a diagonal matrix whose elements can 
be expressed in terms of a scalar function $\lambda^{+}(\omega)$

\begin{equation}
\hat{z}^{+}(\omega) = -t^{2} \left(
\begin{array}{cc}
 f_{2}f_{3} \frac{ \delta^{+}_{0} \delta^{+}_{1}}{ \lambda^{+}_{0} 
\lambda^{+}_{1}}  &   0   \\
  0    &  f_{1}f_{2} \frac{ \delta^{+}_{-1} \delta^{+}_{0}}
{ \lambda^{+}_{0} \lambda^{+}_{1}} 
\end{array} \right) , 
\end{equation}

\noindent
where $\lambda^{+}_{n}= \lambda^{+}(\omega + n\omega_{0}/2)$ and 
$\delta^{+}_{n}= (\lambda^{+}_{n} - g_{n+2})/ t^{2} f^{2}_{n+2}$.
The function $\lambda^{+}(\omega)$ satisfies the following equation

\begin{equation}
\lambda^{+}_{0}= a + b\lambda^{+}_{0} + c\lambda^{+}_{2} + 
d \lambda^{+}_{0} \lambda^{+}_{2} ,
\end{equation}

\noindent
where, taking into account Eq. (17), the coefficients can be 
written as $a=g_{2}+ (t^{2}/W^{2})g_{3}$, $b=t^{2}g_{2}g_{3}$,
$c=-t^{2}[g_{2}g_{3}- (t^{2}/W^{4})]$ and $d=t^{2}[g_{3}+(t^{2}/W^{2})g_{2}]$.
For an arbitrary bias voltage, Eq. (A9) can only be
solved numerically. However, as we show in Appendix B and C
it is possible to obtain analytical solutions in the two special cases 
$\omega_{0} \rightarrow 0$ and $\omega_{0} \rightarrow \infty$.

Once $\hat{z}^{+}(\omega)$ has been determined
one can calculate the coefficients $\hat{T}_1$ and $\hat{T}_{-1}$ 
from Eq. (A7)
                                          
\begin{equation}
\hat{T}_{1} = \frac{-t} {1- t^{2} \lambda^{-}_{2} \lambda^{+}_{-1}} \left(
\begin{array}{cc}
 t^{4} f_{0}f_{1} \delta^{-}_{2} \delta^{+}_{-1}  &  
 t^{2} f_{1} \delta^{+}_{-1} \\ 
 t^{2} f_{0} \delta^{-}_{2}   &  1
\end{array} \right)  
\end{equation}

\begin{equation}
\hat{T}_{-1}(\omega,V)= - \hat{\sigma}_{x} \hat{T}_{1}(\omega,-V) 
\hat{\sigma}_{x} ,
\end{equation}

\noindent
where $ \lambda^{-}_{n}(\omega,V)=\lambda^{+}_{n}(\omega,-V)$ and 
$ \delta^{-}_{n}(\omega,V)=\delta^{+}_{n}(\omega,-V)$. The rest of  
the coefficients
$\hat{T}_{n}$ can be calculated from Eqs. (A5) 

\begin{eqnarray}
\hat{T}_{2n+1} &=& \left[ \prod^{n}_{i=1} 
\hat{z}^{+}(\omega + (i-1) \omega_{0}) \right ] \hat{T}_{1}  , 
\hspace{1cm} n>0 \nonumber \\
\hat{T}_{-n} &=& - \hat{\sigma}_{x} \hat{T}_{n}(\omega,-V) 
\hat{\sigma}_{x} , \hspace{1cm} n>0.
\end{eqnarray}

Finally, the current components, separated into its dissipative and
non-dissipative parts (Eqs. (31) and (32)) can be calculated from the
expressions

\begin{equation}
I_0= - \frac{4e}{h} \int^{\infty}_{-\infty} d \omega
\sum_{n=odd} Re \left\{ Tr( \hat{\sigma}_z \hat{T}^{\dagger}_n(\omega) 
\hat{g}^{+-}_n \hat{T}_n(\omega) \hat{g}^a_0) \right\}
\end{equation}

\begin{equation}
I^D_m= - \frac{4e}{h} \int^{\infty}_{-\infty} d \omega
\sum_{n=odd} Re \left\{ Tr( \hat{\sigma}_z \left[
\hat{T}^{\dagger}_{n+m}(\omega - m\omega_0/2) +
\hat{T}^{\dagger}_{n-m}(\omega + m\omega_0/2) \right]
\hat{g}^{+-}_n \hat{T}_n(\omega) \hat{g}^a_0) \right\}
\end{equation}

\begin{equation}
I^S_m=  \frac{4e}{h} \int^{\infty}_{-\infty} d \omega
\sum_{n=odd} Im \left\{ Tr( \hat{\sigma}_z \left[
\hat{T}^{\dagger}_{n+m}(\omega - m\omega_0/2) -
\hat{T}^{\dagger}_{n-m}(\omega + m\omega_0/2) \right]
\hat{g}^{+-}_n \hat{T}_n(\omega) \hat{g}^a_0) \right\}.
\end{equation}

\section{}

In this Appendix we give details on the evaluation of the excess current
for S-S contacts.
In the limit $eV/\Delta \rightarrow \infty$ only the dc current component 
$I_0$ survives. 
The infinite summation over $n$ in Eq. (A13) can be truncated in this case
neglecting the $n > 1$ terms.
This is justified as the products $f_n f_{n+1}$ are negligible in this
limit leading to a vanishing transfer matrix $\hat{z}^{\pm}(\omega)$.
Physically, this is equivalent to neglecting multiple Andreev processes
for $eV/\Delta \gg 1$.  
Then, Eq. (A13) reduces to

\begin{equation}
I_0 \sim - \frac{4e}{h} \int^{\infty}_{-\infty} d \omega
\sum_{n=-1,1} Re \left\{ Tr( \hat{\sigma}_z \hat{T}^{\dagger}_n(\omega)
\hat{g}^{+-}_n(\omega) \hat{T}_n(\omega) \hat{g}^a_0) \right\} ,
\end{equation} 

\noindent
with

\begin{eqnarray}
\hat{T}_{1} & \sim & \frac{-t} {1- t^{2} \lambda^{-}_{2} \lambda^{+}_{-1}}
\left(
\begin{array}{cc}
 0      &     t^{2} f_{1} \delta^{+}_{-1} \\
 t^{2} f_{0} \delta^{-}_{2}   &  1
\end{array} \right) \nonumber \\
\hat{T}_{-1}(\omega,V) &= &- \hat{\sigma}_{x} \hat{T}_{1}(\omega,-V)
\hat{\sigma}_{x} . \nonumber
\end{eqnarray}

On the other hand, when neglecting contributions of order $\Delta/eV$
Eq. (A9) simply yields $\lambda^+_n \sim
(g_{n+2} + i t^2/W^3)/(1 - i t^2 g_{n+2}/W)$.
We then obtain from Eq. (B1) 
the simple result $I^{SS}_{exc}=
2I^{NN}_{exc}$, for the excess current at zero temperature and any value
of the transmission coefficient.

\section{}

In this Appendix we give the main steps in the analytical 
calculation of the
current components in the limit $eV/\Delta \rightarrow 0$.   

\

{\bf Linear regime ($\eta \gg eV$)}

The small voltage response can be 
straightforwardly derived from Eqs. (A13-A15) by expanding the Fermi functions 
appearing in $\hat{g}^{+-}$ up to first order in $eV$: $n_F(\omega+n
\omega_0/2) \sim n_F(\omega) - (\beta/8) n \omega_0 \mbox{sech}^2(\beta
\omega/2)$ and evaluating the rest of these expressions at $eV = 0$.
The current components can be then
written as

\begin{equation}
I_0=  \frac{2e^2}{h} \beta V \int^{\infty}_{-\infty} d \omega 
\; \mbox{sech}^2(\frac{\beta\omega}{2}) \sum_{n=odd>0} n
Re \left\{ Tr( \hat{\sigma}_z \hat{T}^{\dagger}_n
(\hat{g}^a_0-\hat{g}^r_0) \hat{T}_n \hat{g}^a_0) \right\}
\end{equation}

\begin{equation}
I^D_m=  \frac{2e^2}{h} \beta V \int^{\infty}_{-\infty} d \omega 
\; \mbox{sech}^2(\frac{\beta\omega}{2}) \sum_{n=odd>0} n
Re \left\{ Tr( \hat{\sigma}_z \left[ \hat{T}^{\dagger}_{n+m}
+ \hat{T}^{\dagger}_{n-m} \right]
(\hat{g}^a_0-\hat{g}^r_0) \hat{T}_n \hat{g}^a_0) \right\}
\end{equation}

\begin{equation}
I^S_m=  \frac{8e}{h} \int^{\infty}_{-\infty} d \omega \; n_F(\omega)
\sum_{n=odd>0} Im \left\{ Tr( \hat{\sigma}_z \left[
\hat{T}^{\dagger}_{n+m} -
\hat{T}^{\dagger}_{n-m} \right]
(\hat{g}^a_0-\hat{g}^r_0) \hat{T}_n \hat{g}^a_0) \right\}.
\end{equation}

Thus, the dissipative contribution, $I_D$, goes to zero as $I_D(\phi)
\sim G(\phi) V$, $G(\phi)$ being the phase-dependent linear conductance,
while the supercurrent part, $I_S(\phi)$, tends to a finite value at 
$V=0$.

In the zero voltage limit the coefficients $\hat{T}_{n}$ 
adopt a simple form. The transfer matrix 
$\hat{z}^{\pm}(\omega)$ becomes a scalar function: $\hat{z}^+(\omega) =
\hat{z}^-(\omega) \equiv z(\omega) \hat{I}$; with $z(\omega)= 
-t^2 f \delta^2/ \lambda^2$, where $\lambda^+(\omega)=
\lambda^-(\omega) \equiv \lambda(\omega)$ satisfies the simple quadratic 
equation

\begin{equation}
t^2 g(\omega) \lambda^2(\omega)- (1-\frac{t^2}{W^2}) \lambda(w) 
+g(\omega)=0 .
\end{equation}

\noindent
Finally, the coefficients $\hat{T}_n$ adopt the form

\begin{equation}
\hat{T}_{1}(\omega) = \frac{-t} {1- t^{2} \lambda^2}
\left(
\begin{array}{cc}
 t^{4} f^2 \delta^2       &     t^{2} f \delta \\
 t^{2} f \delta           &           1
\end{array} \right)
\end{equation}

\begin{equation}
\hat{T}_{2n+1}(\omega) = z^n(\omega) \hat{T}_1(\omega), 
\hspace{1cm} n \ge 0.
\end{equation}

Due to these simple recursive relations, the series
appearing in the current components become geometrical series, which can be
summed up without difficulty.
In the weakly damped case, $\eta/ \alpha \Delta 
\ll 1$, these summations lead to analytical expressions for the
dissipative and non-dissipative parts of the current.
By solving Eq. (C4) up to corrections of order $\eta/\alpha \Delta$
one obtains

\begin{equation}
z(\omega)= e^{i \varphi(\omega)} - \frac{4 \omega \eta}{\alpha \Delta^2}
\left[ i + \mbox{cotg}(\varphi(\omega)) \right] ,
\hspace{1cm} \Delta \sqrt{1-\alpha}
\le |\omega| \le \Delta ,
\end{equation}

\noindent
where

\[
\varphi(\omega)= \arcsin \left( \frac{2}{\alpha \Delta^2} 
\sqrt{\Delta^2 - \omega^2} \sqrt{\omega^2 - (1-\alpha) \Delta^2} \right).
\]

The summation of the geometrical series yield

\begin{equation}
I_0=  \frac{2e^2}{h} \beta V \int^{\infty}_{-\infty} d \omega
\; \mbox{sech}^2(\frac{\beta\omega}{2})  
Re \left\{A(\omega) \right \} \frac{1+ |z|^2} {(1-|z|^2)^2} 
\end{equation}

\begin{equation}
I^D_m=  \frac{2e^2}{h} \beta V \int^{\infty}_{-\infty} d \omega
\; \mbox{sech}^2(\frac{\beta\omega}{2}) 
Re \left\{ A(\omega) \left[ (z^m+(z^{*})^m) \frac{1+ |z|^2} {(1-|z|^2)^2}
+ \frac{2mz^m} {1-|z|^2} \right] \right \}
\end{equation}

\begin{equation}
I^S_m=  \frac{8e}{h} \int^{\infty}_{-\infty} d \omega \; n_F(\omega)
Im \left\{ A(\omega) \frac{(z^*)^m - z^m} {1-|z|^2} \right\},
\end{equation}

\noindent
where $A(\omega) \equiv Tr[\hat{\sigma}_z \hat{T}^{\dagger}_1 
(\hat{g}^a-\hat{g}^r) \hat{T}_1 \hat{g}^a]$. It can be noticed that in this
weakly damped limit the integrands
goes like  $1/\eta$ and the energy interval 
$\Delta \sqrt{1-\alpha} \le |\omega| \le \Delta$ give the main 
contribution to the current. 

Finally, when summing up all ac components to obtain the total
dissipative and non-dissipative parts, the current densities become
singular at the condition $\varphi(\omega) = \phi$. This condition 
is satisfied for $\omega = \omega_S = \pm \Delta \sqrt{1 - \alpha
\sin^2(\phi/2)}$ (i.e. at the bound states energy levels) 
leading to   

\begin{equation} 
I_D(\phi) = \frac{e^2 \alpha^2 \Delta^4}{8 \eta h} \beta V 
\int^{\infty}_{-\infty} d \omega \; \mbox{sech}^2(\frac{\beta \omega}{2})
\frac{\sin^2 \varphi(\omega)} {\omega}
Im \left\{ \frac{1}{(\omega- |\omega_S| -i\eta)(\omega+|\omega_S| -i\eta)}
\right\} 
\end{equation}

\begin{equation}
I_S(\phi) = - \frac{2e}{h} \alpha \Delta^2 \sin \phi 
\int^{\infty}_{-\infty} d \omega \; n_F(\omega)
Im \left\{ \frac{1}{(\omega-|\omega_S| -i\eta)(\omega+|\omega_S| -i\eta)}
\right\} .
\end{equation}

These integrals can be straightforwardly evaluated. Eq. (C11) gives the
expression for the phase-dependent linear conductance given in section
IV (Eq. (34)), while Eq. (C12) yields the well known expression for the
supercurrent in a single mode SQPC \cite{joslar,Bennakker}

\begin{equation}
I_S(\phi) = \frac{e \Delta^2 \alpha}{2 \hbar}
\frac{\sin(\phi)}{|\omega_S(\phi)|} 
\tanh \left( \frac{\beta |\omega_S(\phi)|}{2}
\right).
\end{equation}

\

\

\

{\bf Non-linear regime ($\eta \ll eV$)}

We first rewrite Eqs. (A13-A15) for the current components as

\begin{eqnarray}
I_0 &= & \frac{4e}{h} \int^{\infty}_{-\infty} d \omega
\tanh \left( \frac{\beta \omega}{2} \right) \sum_{n=odd>0}
Re \left\{ Tr( \hat{\sigma}_z \hat{T}^{\dagger}_n
(\hat{g}^a_0 - \hat{g}^r_0) \hat{T}_n
\hat{g}^a_{-n}) \right\} \nonumber \\
I^D_m &=&  \frac{4e}{h} \int^{\infty}_{-\infty} d \omega
\tanh \left( \frac{\beta \omega}{2} \right) \sum_{n=odd>0}
Re \left\{ Tr( \hat{\sigma}_z \left[
\hat{T}^{\dagger}_{n+m} \hat{T}^{\dagger}_{n-m} \right]
(\hat{g}^a_0 - \hat{g}^r_0) \hat{T}_n
\hat{g}^a_{-n}) \right\} \nonumber \\
I^S_m &=& - \frac{4e}{h} \int^{\infty}_{-\infty} d \omega
\tanh \left( \frac{\beta \omega}{2} \right) \sum_{n=odd>0}
Im \left\{ Tr( \hat{\sigma}_z \left[
\hat{T}^{\dagger}_{n+m} -
\hat{T}^{\dagger}_{n-m} \right]
(\hat{g}^a_0 - \hat{g}^r_0) \hat{T}_n
\hat{g}^a_{-n}) \right\} ,
\end{eqnarray}
where a rigid shift of $n \omega_0/2$ in the energy
arguments of the different
$\hat{T}_n$ with respect to the ones appearing in Eq. (A13-A15)  
has been introduced. 

In the limit $eV \rightarrow 0$, the solution of Eq. (A9) is 
$\lambda^+_n = \lambda[\omega + (n+2)\omega_0/2]$, where
$\lambda(\omega)$ satisfies the quadratic Eq. (C4).  
The coefficients $\hat{T}_n$ ($n > 0$)
can then be generated starting form $\hat{T}_1$ and
using the transfer matrix $\hat{z}^+(\omega)$. These quantities 
are obtained from Eqs. (A8) and (A10) making use of the 
$eV \rightarrow 0$ solution for $\lambda^+_n$.

The resulting expressions simplify considerably in the ballistic limit
where $\lambda^+_n = i/t$. For energies inside the gap one obtains

\begin{equation}
\hat{T}_n(\omega) = -\frac{t}{2} \prod^{n-1}_{j=1} e^{i \alpha_j}
\left(
\begin{array}{cc}
 e^{i (\alpha_0 + \alpha_n)}       &     e^{i \alpha_n}  \\
 e^{i \alpha_0}           &           1
\end{array} \right) ,
\end{equation}
where $\alpha_n = \mbox{arccos}[(\omega + neV)/\Delta]$. As discussed in
Ref. \cite{Averin}, when written as in Eq. (C14), 
the main contribution to the current in this limit comes from a small
energy range around the gap edges. Evaluation of these 
integrals leads directly to Eq. (35).

\begin{figure}
\caption{Schematic representations of the two different situations
discussed in Section II.}
\end{figure}

\begin{figure}
\caption{The dc current-voltage characteristic of a S-S contact for
different values of the normal transmission at zero temperature.} 
\end{figure}

\begin{figure}
\caption{The first three ac components of the dissipative current for
different values of the normal transmission at zero temperature.}
\end{figure}

\begin{figure}
\caption{Same as Fig. 4 for the non-dissipative current.}
\end{figure}

\begin{figure}
\caption{Current density for the dc component $I_0$ within the linear
regime discussed in section IV.C. Cases (a), (b) and (c) correspond to
transmission values $\alpha = 1$, 0.65 and 0.04 respectively. In all
cases the full line corresponds to $\eta/\Delta = 1/10$, the dotted line
to $\eta/\Delta=1/25$ and the broken line to $\eta/\Delta=1/100$. The
thermal factor $\mbox{sech}^2(\beta \omega/2)$ (see Eq. (C8))
has been extracted from the current density.}
\end{figure}

\begin{figure}
\caption{Behavior of the first non-dissipative (a) and dissipative (b)
ac components in the very small voltage range close to ballistic
conditions. These results have been obtained for negligible $\eta$.}
\end{figure}


\begin{references}
\bibitem{comment2} A finite order perturbative calculation in $t$ yields
a conductance which diverges at $eV = \Delta$ due to the divergencies of
the uncoupled BCS Green functions (Eq. (7)).
\end{references}
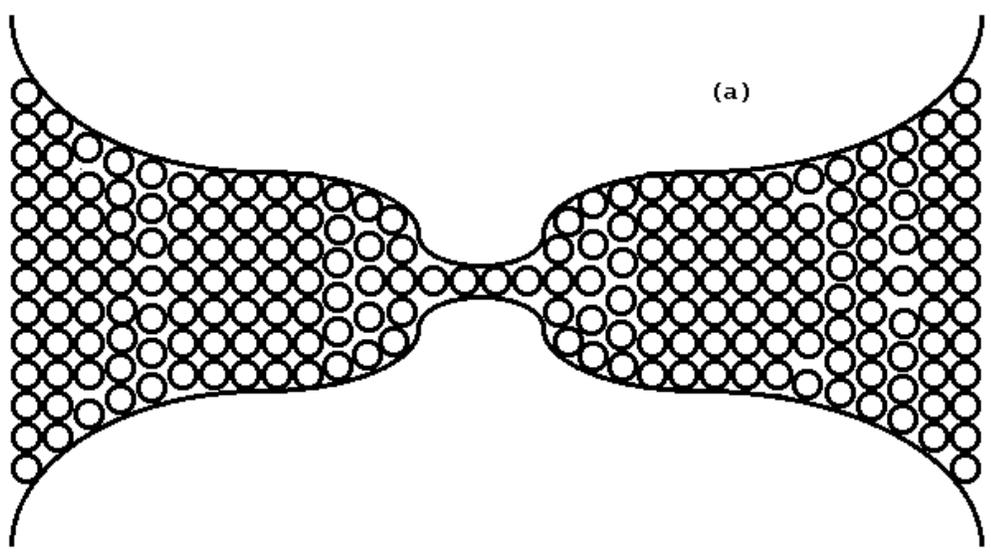
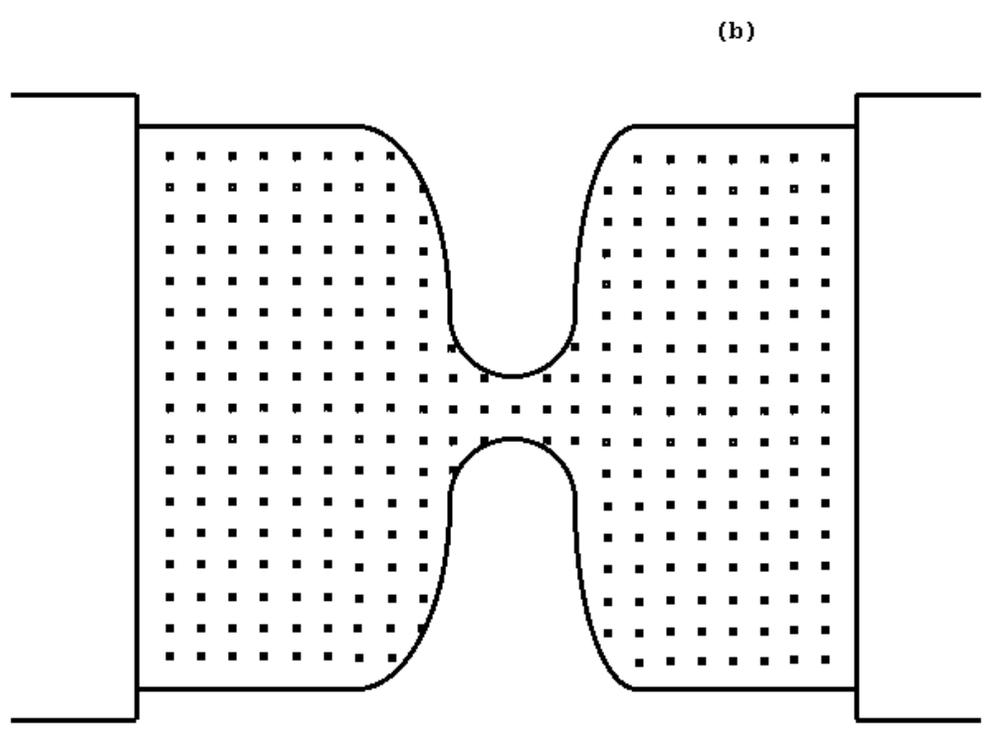
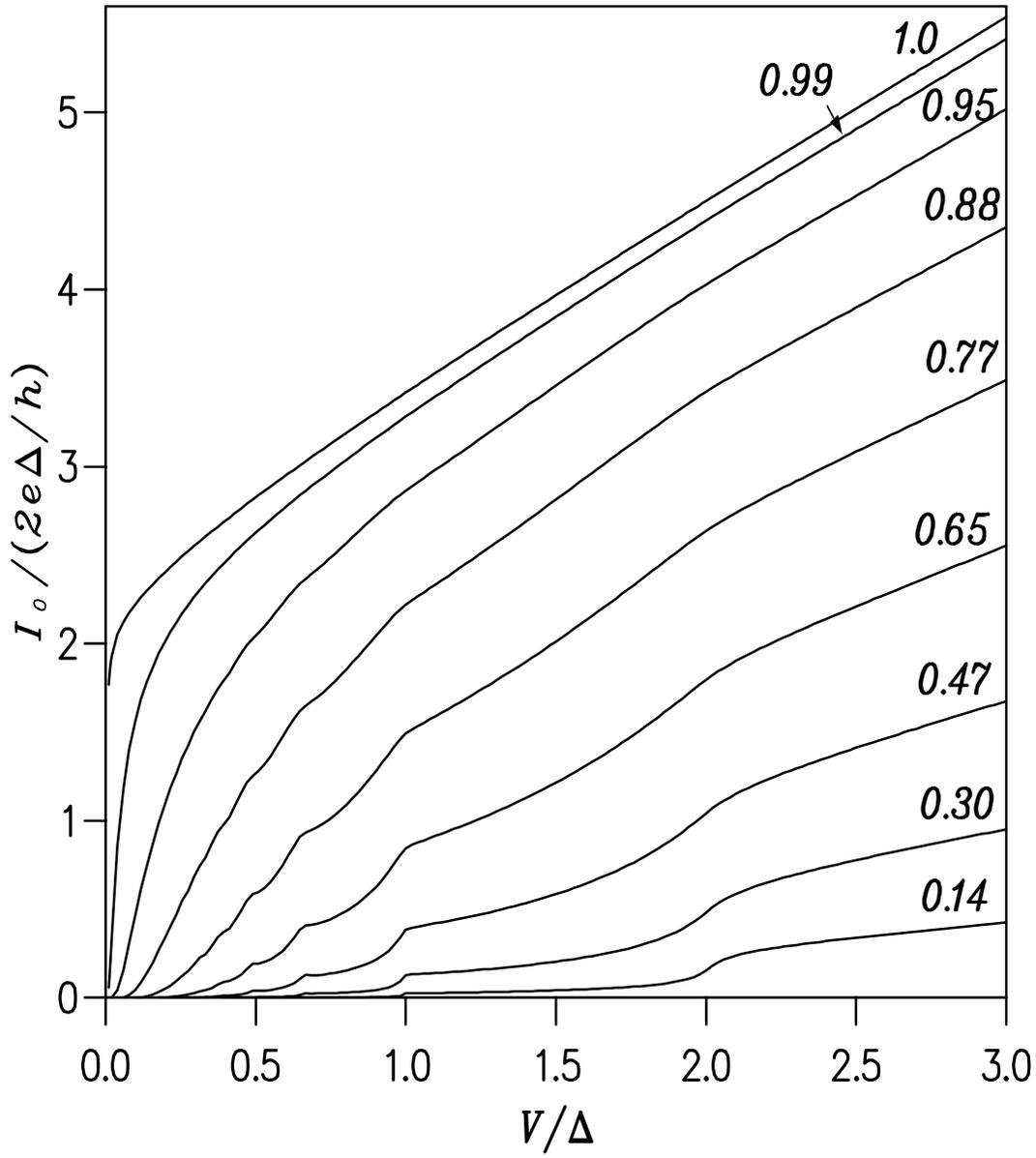
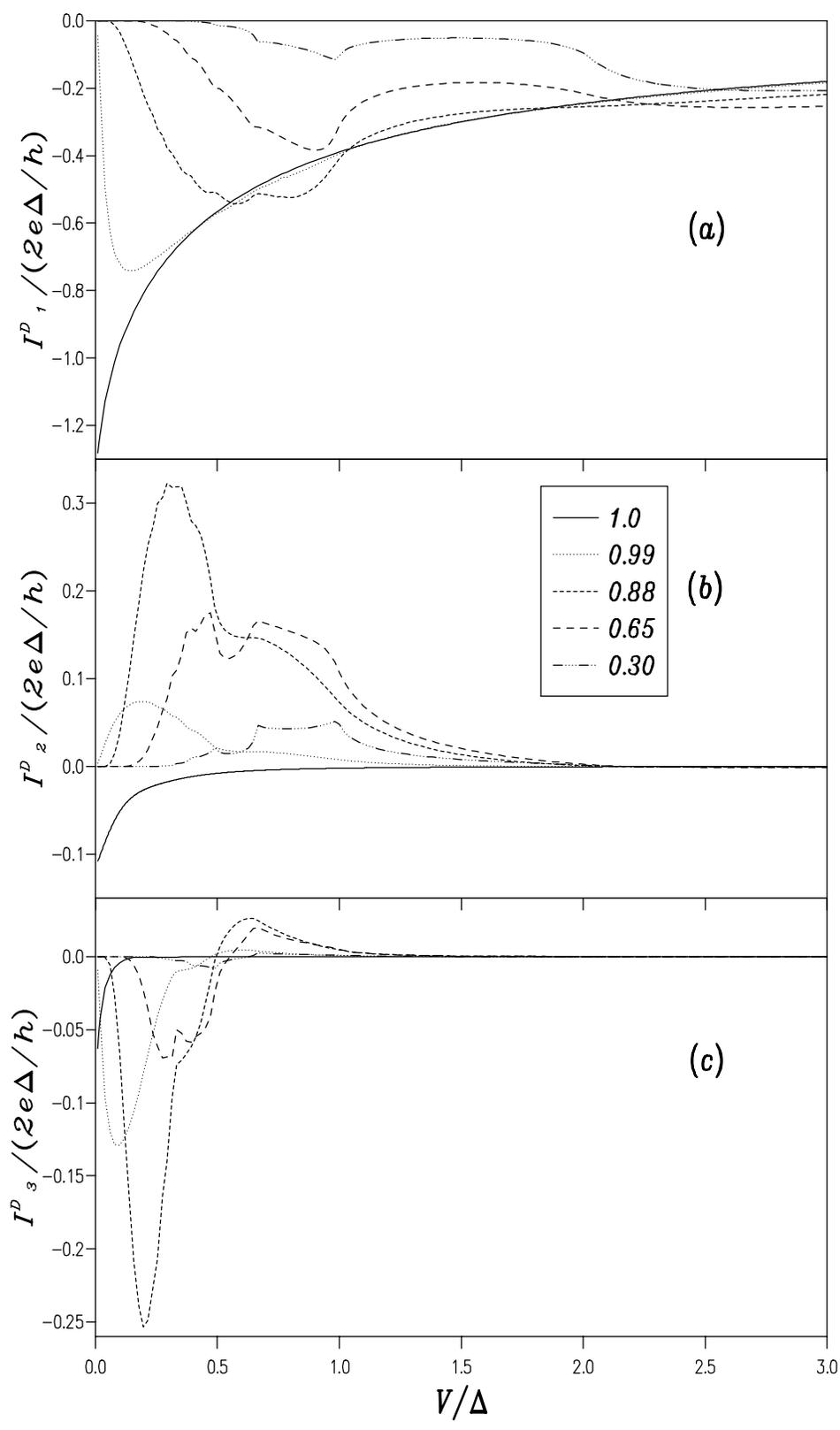
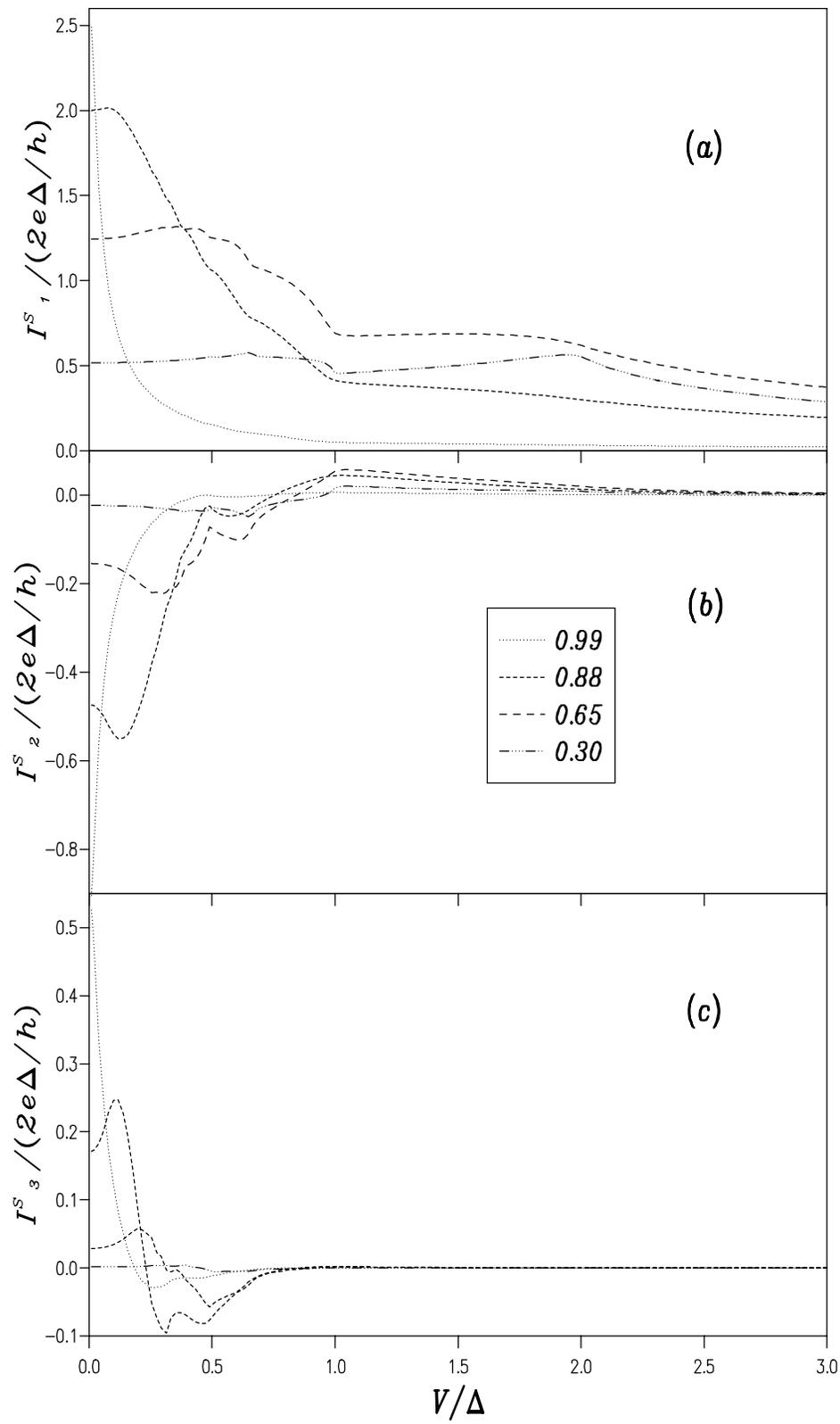
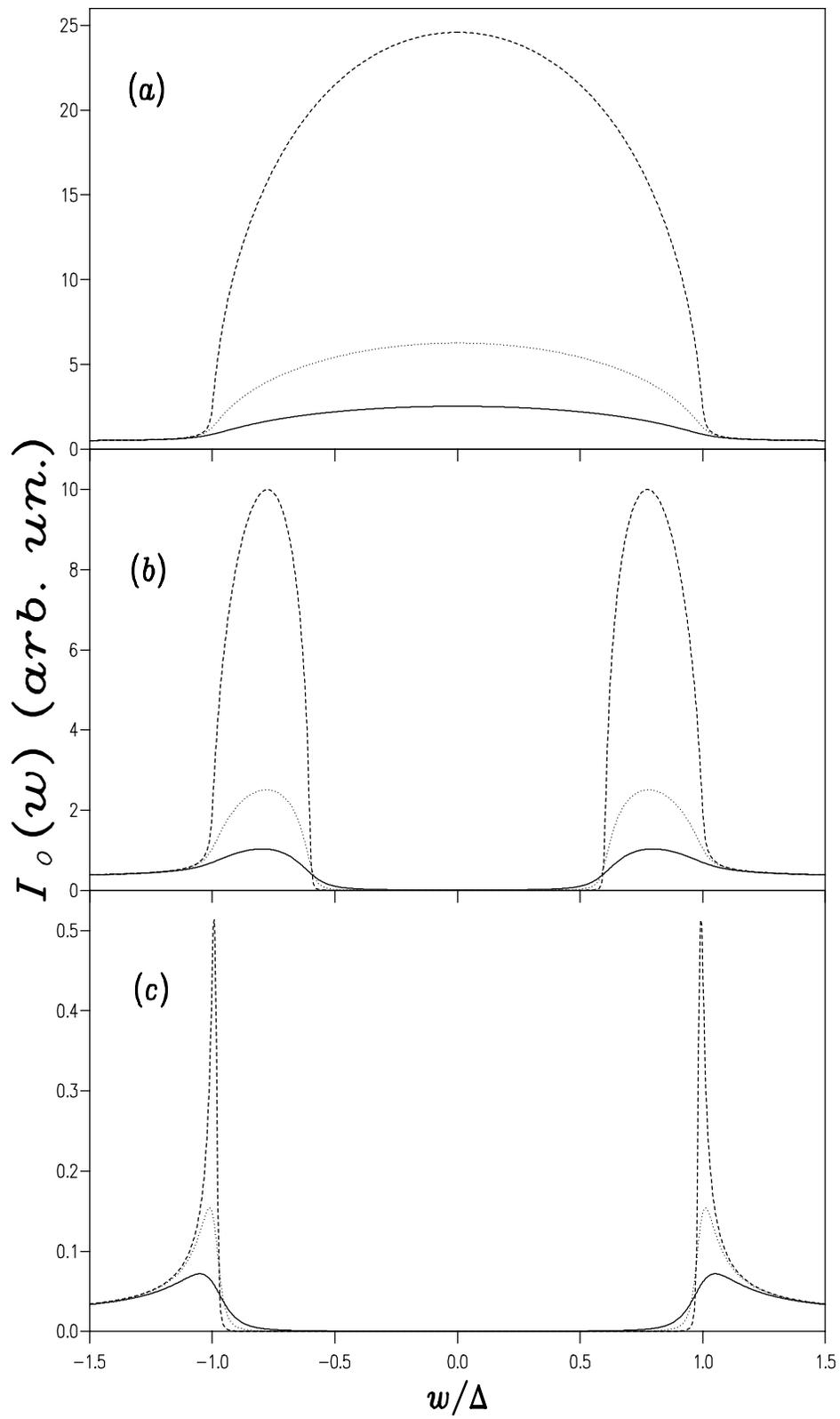
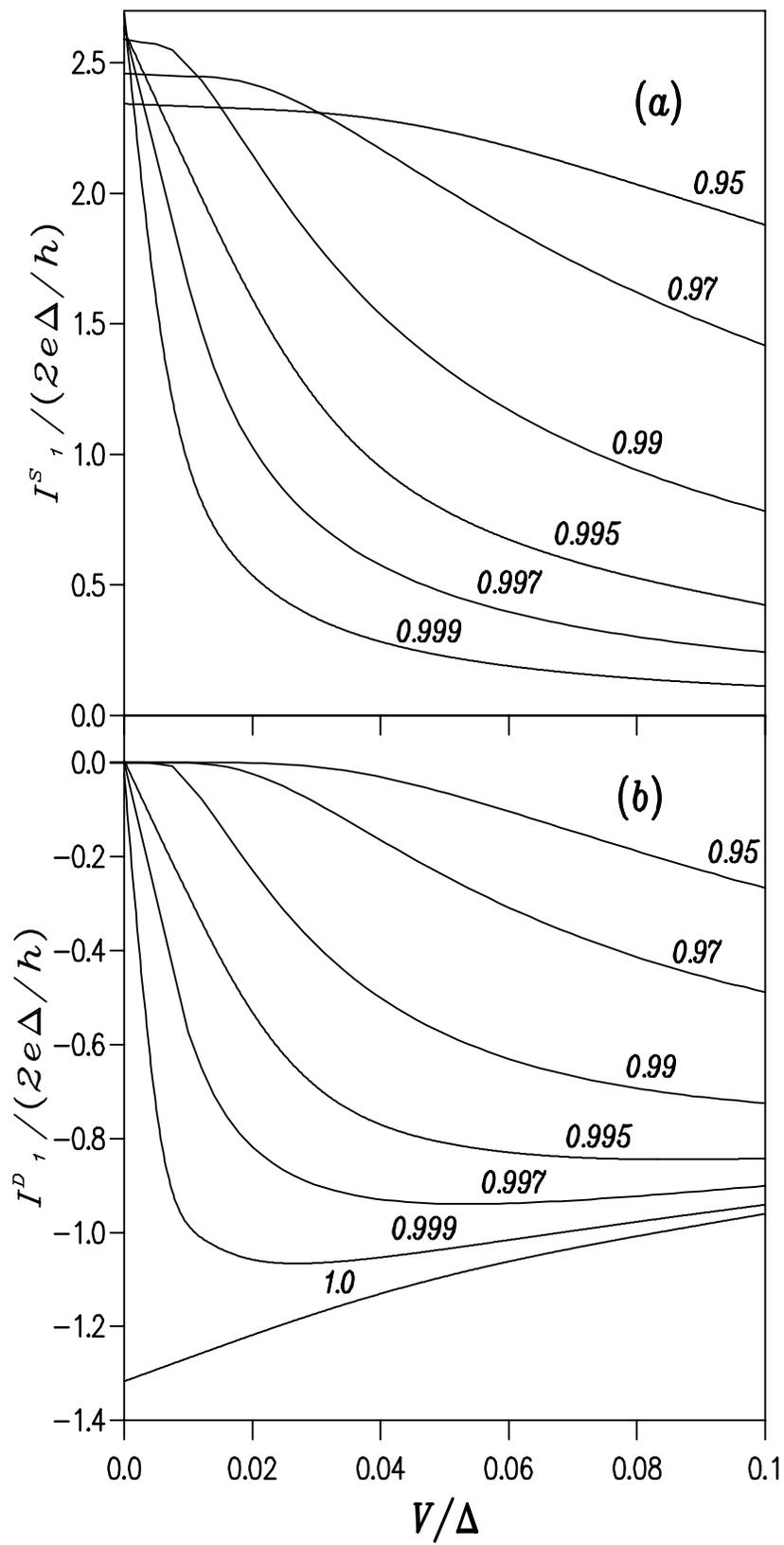
\end{document}